\documentstyle[twocolumn,aps]{revtex}
\newcommand{\be}{\begin{equation}}
\newcommand{\ee}{\end{equation}}
\newcommand{\bea}{\begin{eqnarray}}
\newcommand{\eea}{\end{eqnarray}}

\begin{document}

\draft

\title{Detection of vorticity in Bose-Einstein condensed gases by 
matter-wave interference}
\author{Eric L. Bolda and Dan F. Walls}
\address{Department of Physics, University of Auckland, Private Bag 92019, 
\\ Auckland,
New Zealand}
\date{\today}
\maketitle
\begin{abstract}
A phase-slip in the fringes of an interference pattern is an 
unmistakable characteristic of vorticity. We show dramatic 
two-dimensional simulations of interference between expanding condensate 
clouds with and without vorticity.
In this way, 
vortices may be detected even when the core itself cannot be resolved.
\end{abstract}
\vspace{1 ex}

%\section{Introduction}

Bose-Einstein condensation of a dilute gas has now been accomplished 
for three different alkali atoms by a number of research groups 
\cite{BECexpt}.  Many of the properties of the gas which were 
calculated from mean-field theories, such as condensate density, 
collective excitation spectra, and the speed of sound have also been 
found to agree well with these theories.  
One of the untested predictions of mean-field theory is the 
quantization of circulation of the fluid, leading to quantized vortex 
states.  In this Letter we show how such a state may be detected by using 
matter-wave interference.  This method may allow us to follow the dynamics 
of states with vorticity as well.

Several ways of {\em creating} vortices in a Bose-Einstein 
condensate (BEC) have been suggested.  Initial experimental 
attempts relying on a rotating, off-resonance laser beam were 
inconclusive \cite{Ketterle98pr}.  Another idea is that vortices 
may appear spontaneously out of fluctuations, as a result of 
rapid cooling through the critical temperature \cite{Anglin98a}.  
Both of the foregoing methods are analogous to techniques in more 
traditional superfluids such as liquid $^{4}He$.  Unlike those 
systems, the internal states of the alkali atoms provide a 
convenient ``handle'' for the optical creation and detection of 
vortices.  Starting with the trap ground state, vortex states can 
be reached via a Raman transition quickly 
\cite{Bolda98b,Marzlin97} or adiabatically \cite{Dum98}.

Here we focus on the {\em detection} of states with vorticity or 
non-zero winding number.  Indirect detection methods have 
previously been suggested \cite{indirect}, as has detection of 
the spatially-dependent phase in an ionization scheme 
\cite{Goldstein98}.  The interference of condensates released 
from a double-well trap \cite{Andrews97} was the first evidence 
of their phase coherence.  At a vortex core, the phase will 
always have a singularity which is independent of atomic and trap 
parameters.  Since there is coherence over most of the 
condensates, we suggest that the phase slip occuring at a vortex 
core be used to detect it.  When colliding a condensate in the 
ground state with a (presumed) vortex state, the vorticity is 
measured by counting the number of fringes which are ``skipped'' 
as they cross the core.

%  This is a global effect and does not 
% require vorticity to be calculated using atomic and trap parameters 
% which may not be accurately known (e. g. number of atoms).
% is dislocations really the right term? ``singularities''?
% or ``phase slip''

%\section{Vorticity in Bose-Einstein condensates}
We briefly review the theory of the BEC far below the critical temperature 
beginning from the Gross-Pitaevski (GP) equation for the condensate 
wavefunction $\psi({\mathbf x},t)$,
\bea
i \frac{\partial \psi}{\partial t} & = & -\frac{1}{2} \nabla^{2} \psi + V({\mathbf 
x}) \psi + \frac{u}{2} |\psi|^{2} \psi.
\eea
In the case of axisymmetric harmonic traps, the external potential is
\be
V(r,z) = \frac{1}{2}(r^{2} + \lambda^{2} z^{2})
\ee
Here the time is scaled in units of the reciprocal trap frequency, 
$\frac{1}{\omega_{\bot}}$, and all lengths are scaled in units of the transverse trap width,
\bea
a_\bot	& =	& \sqrt{\frac{\hbar}{M \omega_{\bot}}} \\
\lambda & = & \frac{\omega_{\bot}}{\omega_{z}}
\eea
so that the dimensionless interaction strength is
\bea
u & = & \frac{8 \pi a N}{a_\bot}
\eea
where $N$ is the number of condensate atoms, and $a$ is the 
$s$-wave scattering length.  Since we are interested primarily in the 
two-dimensional dynamics, we 
assume that the $z$ dimension is thin enough that all dynamics in that 
direction can be neglected ($\lambda \gg 1$).

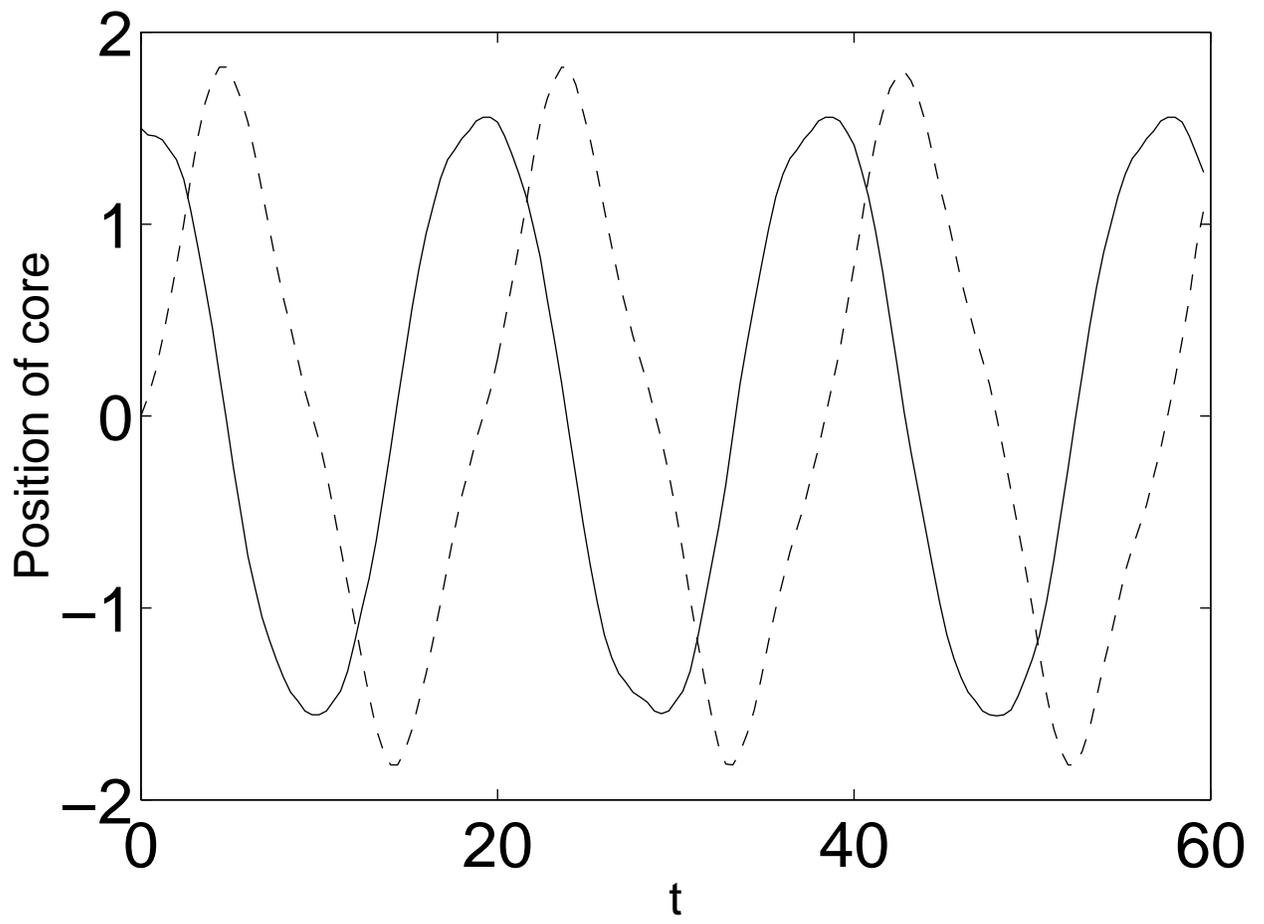
\begin{figure}
\caption{Position of a precessing core in a generic vorticity state.  {\em Dashed 
curve}, $x$; {\em solid curve}, $y$ coordinate. The initial state was prepared 
by imaginary time evolution subject to constrained normalization and 
phase. Interaction strength $u = 300$.}
\label{core_vs_t.ps}
\end{figure}

We distinguish between {\em  pure vortex} states and generic states with 
{\em vorticity}.  The latter are any states for which there exists a closed 
curve $C$ such that the circulation (in unscaled variables) obeys
\bea
\Gamma = \oint_{C} {\mathbf v} \cdot \, d {\mathbf x} = \frac{h m}{M} 
\neq 0.
\eea
Owing to the single-valuedness of the wavefunction, $m$ must be an integer.  
The structure near the vortex core is always similar: for 
cylindrical coordinates $(\rho, \phi)$ centered at $(x_c,y_c)$, the density goes 
from $0$ to $n_{0}$ within a radius of the order of the healing 
length
\be
\xi = \frac{|m|}{\sqrt8 \pi n_{0} a},
\ee
and the phase is 
\be
\phi = m \varphi = m \arctan \frac{y-y_c}{x-x_c}.
\label{phieqn}
\ee
Circulation is conserved around any closed curve flowing with the 
condensate (Kelvin's theorem).  On the other hand, these states are not 
stationary; our numerical solution of the GP equation shows precession 
around the trap center with a period depending on the distance from it to 
the core and on $u$.  For example, when $u = 300$, a core $1.5$ from the 
center precesses with a period $\approx 19$ (Fig.  \ref{core_vs_t.ps}).  
This appears to violate Ehrenfest's theorem; however, the precession is 
accompanied by considerable density excitations so that the center of mass 
still obeys classical equations of motion.  From this we may expect that 
the dynamics of generic vorticity states depends not only on the atomic 
collisions but also damping rates of excitations.

In the case of axisymmetric traps, a subset of the {\em 
vorticity} states are {\em pure vortex} states of the form
\bea
\psi(r, z, \varphi) = f(r,z) e^{i m \varphi}
\eea
These states are stationary and are eigenstates of azimuthal angular momentum 
$L_{z} = \hbar m$.
%The pure vortex states are found by substituting $\psi(r,  \varphi) = f(r) 
%\exp(i m \varphi)$ into the GP equation. Numerically, one 
% evolves $\psi$ in imaginary time subject to the constraints of 
% normalization and $\arg \psi =  m \varphi$.
The metastability of these states is still controversial 
\cite{Svidzinsky98,Zambelli98}.
%For the 
%purpose of this paper, we assume that vorticity does not decay before 
%the condensates have expanded and interfered.
However, one decay scenario is that pure vortex states turn into 
vorticity states with the core off-center \cite{Rokhsar97}; 
the possible observation of this 
process is one of the motivations for our detection scheme.

%\section{Simulation of interference}

% Expansion of the core does not change the circulation around it.
% Circulation is conserved along a curve flowing with the fluid
% (Kelvin s theorem).
For interference between a generic vorticity state and a ground state 
moving towards each other with relative velocity $2k$ in the 
$x$-direction, in the core region the wavefunction is approximated by
\bea
\psi(\mathbf{x}) & \simeq & \sqrt{n_{1}}e^{ikx} + \sqrt{n_{2}} 
e^{(-ikx+im \varphi)}
\eea
resulting in a density
\bea
n(\mathbf{x}) & \simeq & n_{1} + n_{2} + 2 \sqrt{n_{1} 
n_{2}} \cos({2 kx - m \varphi})
\eea
This density displays a series of fringes, with $|m|+1$ fringes merging 
into 1 at the origin, which is the location of the vortex core.  Such 
patterns have been used to detect vortices in nonlinear optics 
\cite{Swartzlander92}.  The momentum $k$ may be acquired from free 
expansion of the condensates or by moving the traps.

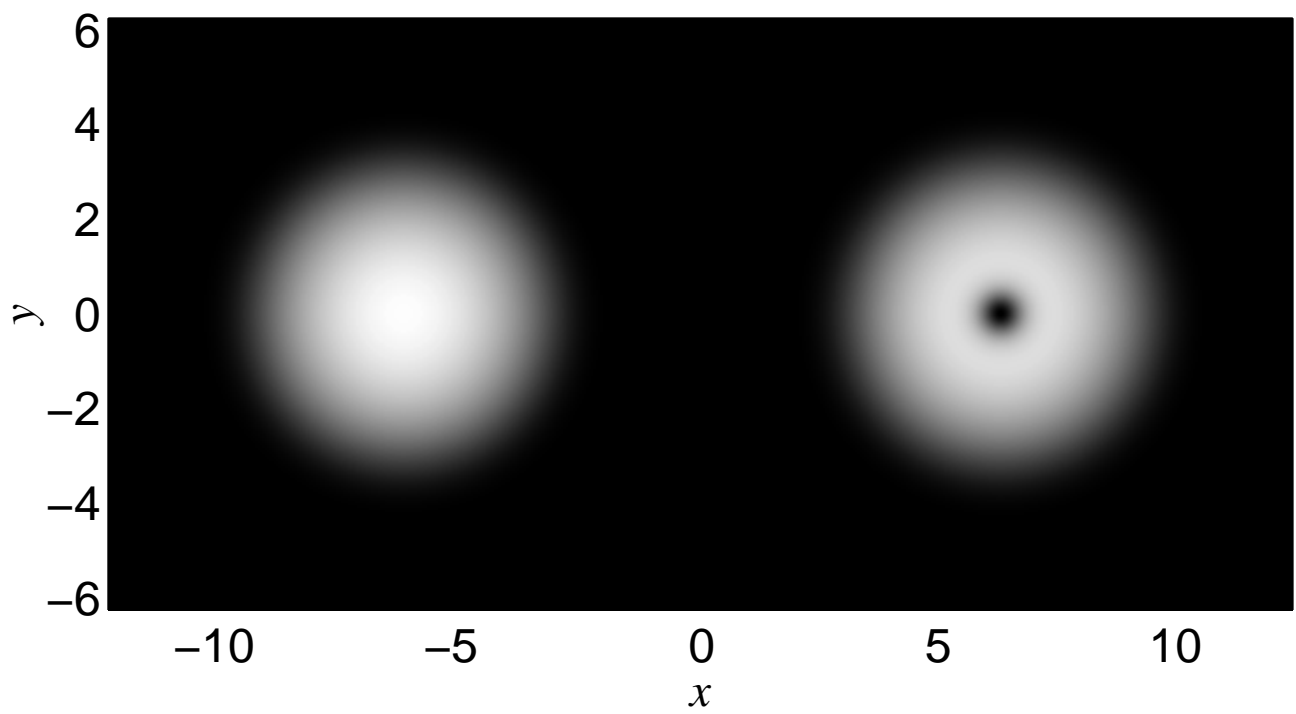
\begin{figure}
\caption{Initial density of two-trap BEC for interference. Right condensate is in the 
$m=-1$ vortex state, left condensate is in the trap ground state. 
Interaction strength $u = 300$.}
\label{k0figinit.ps}
\end{figure}

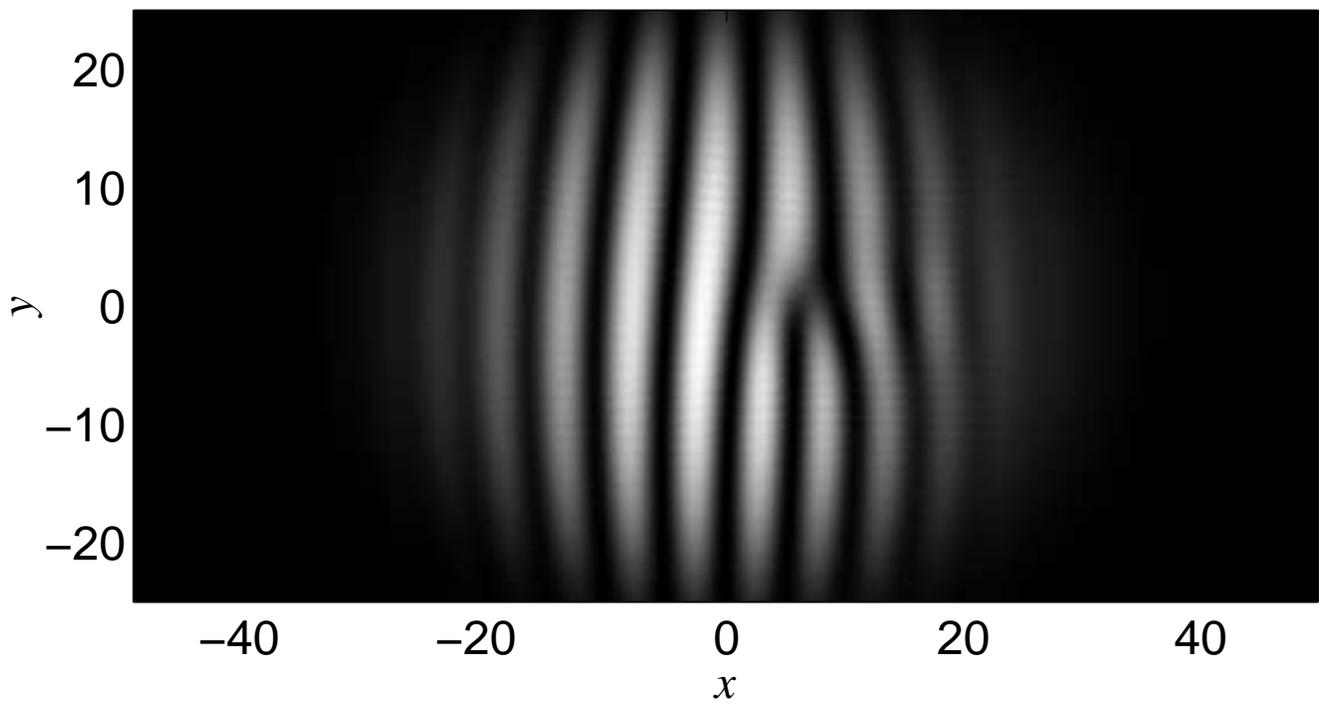
\begin{figure}
\caption{Density at $t = 10$ after the traps of Fig. \ref{k0figinit.ps} have been 
switched off.  The original vortex is clearly evident from the missing fringe 
near $(8,0)$.}
\label{k0figfin.ps}
\end{figure}

We have computed several interference patterns between states 
with and without vorticity, by numerically solving the GP 
equation in two dimensions.  Our initial configuration consists 
of a pair of trapped condensate clouds with equal numbers of 
atoms.  It is not necessary that either the actual number or the 
relative phase between the clouds be known.  In Fig.
\ref{k0figinit.ps}, the cloud centered at $(6.25, 0)$ begins in the 
$m=-1$ pure vortex state while that at $(-6.25, 0)$ is the trap ground 
state. The traps are switched off at $t=0$. 
 Up until $t \simeq 1.5$ the two clouds expand independently.  Then,
 as in the first interference experiment 
\cite{Andrews97}, the ``pulsed'' sources lead to straight fringes.  
However, as the core region of the upper condensate approaches the 
lower condensate, the difference in velocity between the upper and 
lower sides of the vortex tilts the fringes from the vertical.  Finally, 
as the fringes reach the position of the core, the phase slip in fringes appears 
clearly (Fig. \ref{k0figfin.ps}).  We note that the core is smaller 
than the fringe spacing in this example. 

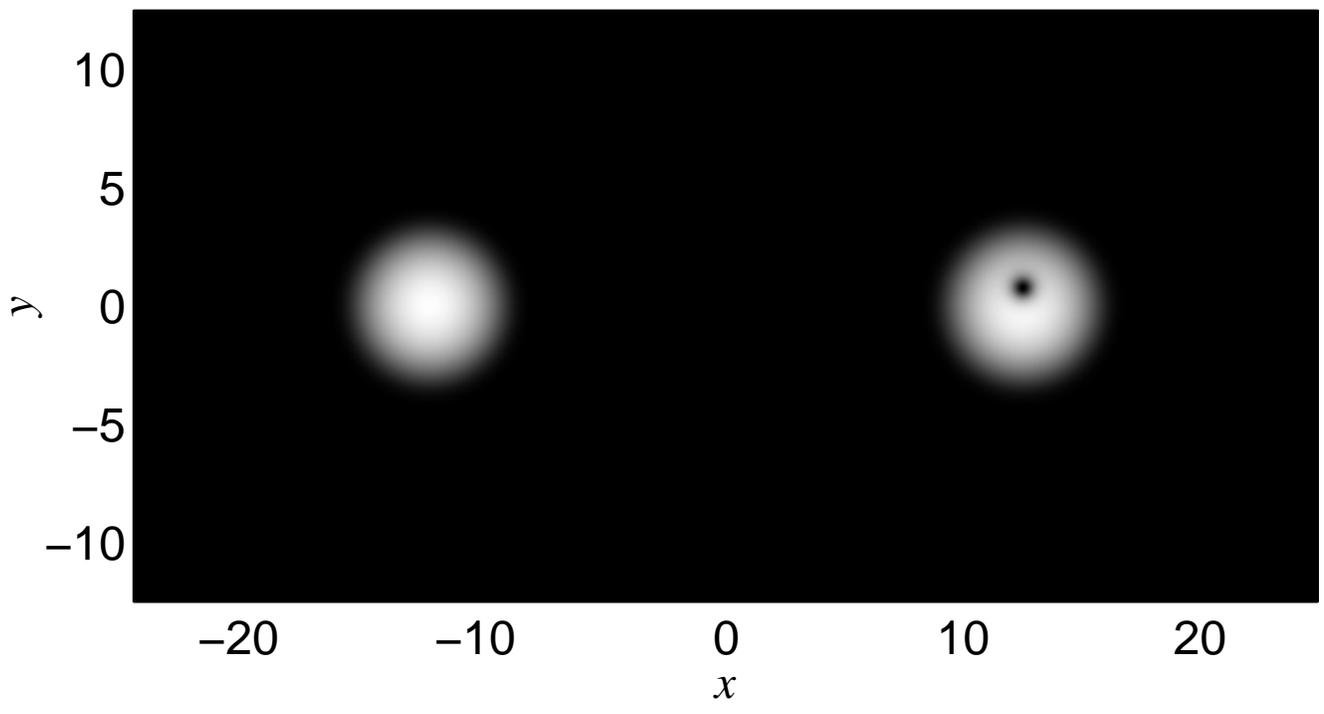
\begin{figure}
\caption{Initial density for two condensates moving towards each 
other.  Left cloud is the ground state, right cloud has an $m=-1$ vortex 
core at $(-12.5, 0.75)$. Interaction strength is $u = 300$.}
\label{k4fig1.ps}
\end{figure}

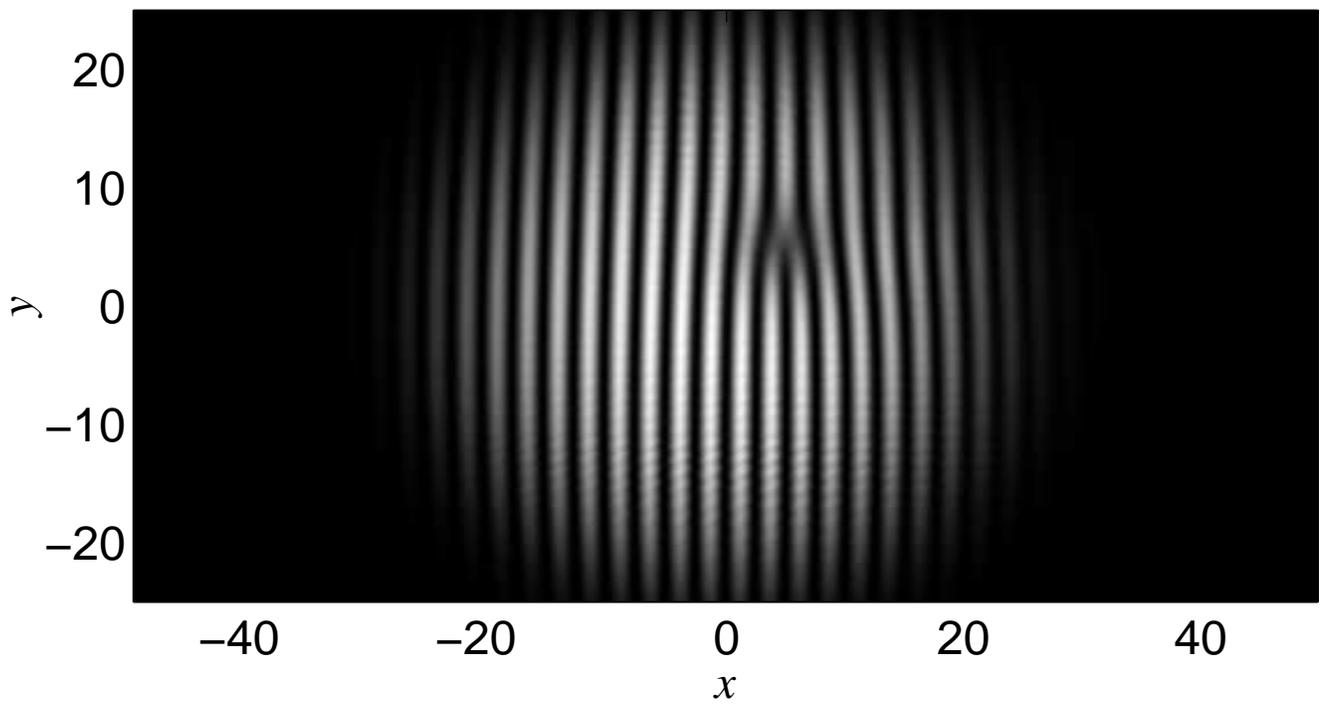
\begin{figure}
\caption{Density at $t = 10$ after the traps of Fig. \ref{k4fig1.ps}
moving at $k = \pm 4$ have been switched off.  The position of the phase 
slip is still above the $x$-axis, and has precessed less than $1/8$ of the 
way around the cloud.}
\label{k4fig2.ps}
\end{figure}

In the second configuration, shown in Fig. \ref{k4fig1.ps}, 
the cloud on the right is replaced by an $m=-1$ generic vorticity state 
with the trap center placed at $(-12.5, 0)$ and the core at $(x_c = 12.5, y_c = 0.75)$.
  This state was produced by 
evolving the wavefunction in imaginary time, subject to the 
constraints of normalization and that the phase be given by 
Eq. (\ref{phieqn}).  The cloud on the left is 
the ground state centered at $(12.5, 0 )$. Also, the condensates are 
now launched towards each other so the wavefunction is multiplied 
by $\exp (ikx)$ for $x<0$ and by $\exp (-ikx)$ for $x>0$, with 
$k = 4$.  Since the vortex is off-center, it begins to precess 
before the two clouds collide.  In this example, the period of 
this precession is much longer than the time until the 
interference pattern reaches the core.  Thus, in Fig. 
\ref{k4fig2.ps} the position of the 
phase-slip above the $x$-axis indicates the initial position of 
the vortex.  This suggests that precession of a core may be observed by 
repeatedly creating an off-center vortex, allowing it to 
evolve in the trap for successively longer times, 
and then interfering it with a ground state.

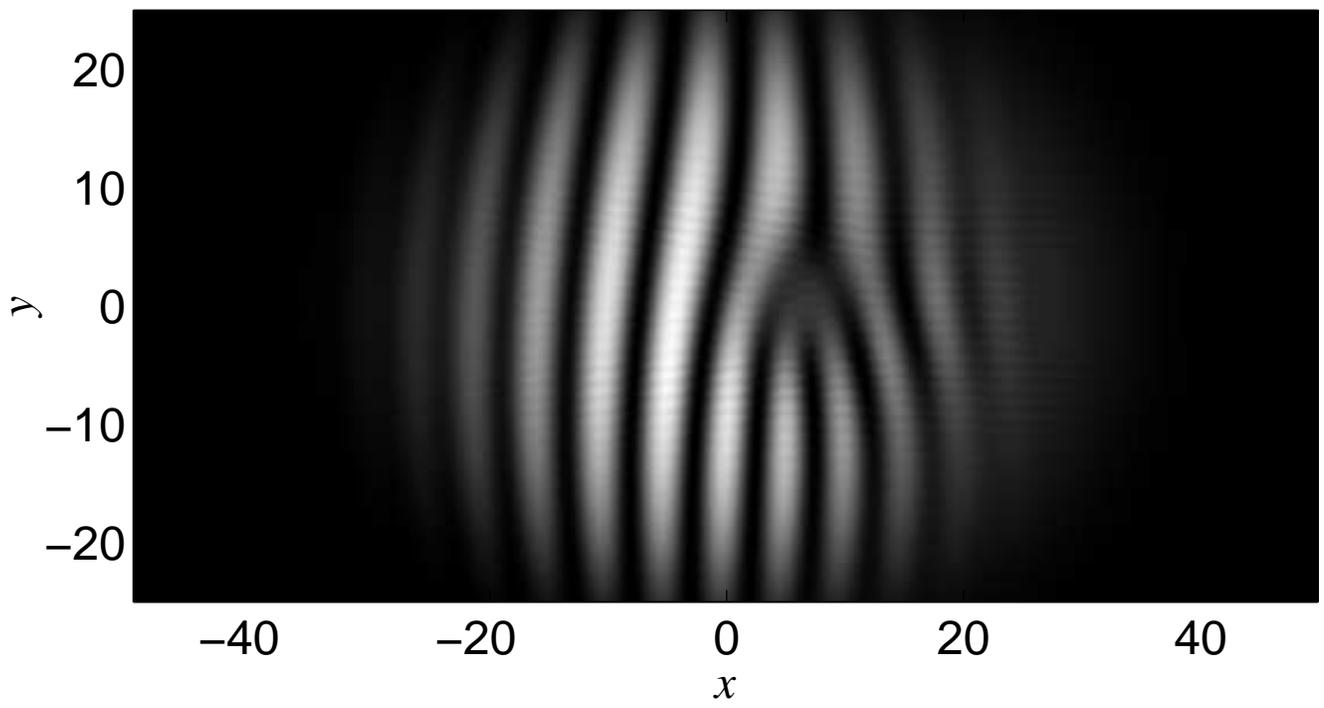
\begin{figure}
\caption{Density at $t = 10$ after free expansion of two trapped 
condensates, with right cloud starting in the $m=2$ vortex state, 
left cloud in the trap ground state. Interaction strength $u = 300$.}
\label{psiexpan_m2_.ps}
\end{figure}

We have also examined the case of vortex states with $|m|>1$.  
Any non-azimuthally symmetric perturbation is expected to break these 
into a collection of 
$m = \pm 1$ generic vortices. 
 Fig. \ref{psiexpan_m2_.ps} is the pattern resulting from freely expanding clouds at 
 $(-6.25,0)$, $m = 0$, and  $(6.25, 0 )$, $m = 2$.  In this case 
 two fringes are skipped; however it is not possible to 
 determine from the density alone that in fact we have two nearby $m=1$ phase 
 slips.

%\section{Conclusion}
In an actual experiment, the core region itself may be obscured 
for several reasons.  First, if the local density is high enough, 
the core radius $\xi$ will be less than the imaging wavelength 
and may not be resolved.  Second, light traveling along the 
$z$-direction will always pass through regions of non-zero density if the 
core is not exactly parallel to the laser beam.  (This problem 
can be avoided by optically pumping atoms in thin sheets 
perpendicular to the imaging beam \cite{Andrews97}).  Finally, 
the position of the core is subject to thermal and zero-point fluctuations 
of the order of its radius \cite{Barenghi96}.  We note that detection of the fringes 
avoids all of these problems.  The same method may also be used to measure 
non-zero winding numbers in toroidal traps.

We have assumed that  damping takes place on longer timescales 
than the dynamics seen in our simulations, which run up to a few 
trap periods.
As mentioned above, to accurately follow the motion of generic vorticity states 
 requires that we go beyond the GP equation to include 
thermal effects.  We are continuing research in this direction.

%Acknowledgments
This research was made possible by the Marsden Fund of the Royal 
Society of New Zealand and The University of Auckland Research 
Fund.  M. J. Dunstan provided valuable assistance with the 
computer coding.

%% 
 % \cite{BECexpt}
 % \cite{Ketterle98pr} % laser ``paddle'' didn't work (ITP conference)
 % \cite{Dalfovo96}	% ground and vortex	wavefunctions
 % \cite{Lundh97} %	estimate of	kinetic	energy etc.
 % \cite{Andrews97}	% interference expt.
 % \cite{Bolda98b}
 % \cite{Marzlin97}
 % \cite{Marzlin98a,Marzlin98b}	% rotation in anharmonic trap 
 % \cite{Dum98}	 % adiabatic creation of vortex	
 % \cite{Anglin98a}
 % 
 % \cite{Mueller98}	% decay	of current in torus	
 % \cite{Rokhsar97}	% decay	of vortex in harmonic trap
 % 
 % \cite{Stamperkurn98}	% dipole trap expt.
 % \cite{Svidzinsky98} % normal	modes of vortex	at T=0
 % \cite{Zambelli98} % frequency shift of oscillations using sum rules
 % \cite{Lundh98} %	detection by expanding clouds, anisotropy effects
 % \cite{Bolda98un}	% increase of thermal cloud	after vortex decay
 % \cite{Swartzlander92} % detection of	vortex in nonlinear	optics
 % \cite{Barenghi96} % helical vortex waves
 % \cite{Wallis97b}	% interference theory, compared	with experiment
 % \cite{Goldstein98} %	vortex msmt	using g2 (ionization scheme)
 %%

\bibliography{erics,detectv}

\begin{thebibliography}{10}

\bibitem{BECexpt}
M.~H. Anderson {\it et~al.}, Science {\bf 269}, 198 (1995); K.~B. Davis {\it
  et~al.}, Phys. Rev. Lett. {\bf 75}, 3969 (1995); C.~C. Bradley {\it et~al.},
  {\it ibid}, 1687 (1995).

\bibitem{Ketterle98pr}
W. Ketterle, private communication.

\bibitem{Anglin98a}
J.~R. Anglin and W.~H. Zurek,   (1998), preprint quant-ph/9804035.

\bibitem{Bolda98b}
E.~L. Bolda and D.~F. Walls, Physics Letters A %{\bf ??},  ????  (1998)
, to be published.

\bibitem{Marzlin97}
K.-P. Marzlin, W. Zhang, and E.~M. Wright, Phys. Rev. Lett. {\bf 79},  4728
  (1997).

\bibitem{Dum98}
R. Dum, J.~I. Cirac, M. Lewenstein, and P. Zoller, Phys. Rev. Lett. {\bf 80},
  2972  (1998).

\bibitem{indirect}
Indirect evidence for vortex states could come from frequency shifts and
  breaking of degeneracy in the excitation spectra
  \cite{Svidzinsky98,Zambelli98}. Of course those detection methods would
  require knowledge of the trap parameters, number of atoms and the scattering
  length, which may not be accurately known. We note that there is not complete
  agreement on the nature of the excitation spectrum; this is also related to
  the question of the stability of the vortex \cite{Rokhsar97}.

\bibitem{Goldstein98}
E.~V. Goldstein, E.~M. Wright, and P. Meystre, Phys. Rev. A {\bf 58},  576
  (1998).

\bibitem{Andrews97}
M.~R. Andrews {\it et~al.}, Science {\bf 275},  637  (1997).

\bibitem{Svidzinsky98}
A. Svidzinsky and A.~L. Fetter,   (1998), preprint cond-mat 9803181.

\bibitem{Zambelli98}
F. Zambelli and S. Stringari,   (1998), preprint cond-mat/9805068.

\bibitem{Rokhsar97}
D.~S. Rokhsar, Phys. Rev. Lett. {\bf 79},  2164  (1997).

\bibitem{Swartzlander92}
G.~A. Swartzlander and C.~T. Law, Phys. Rev. Lett. {\bf 69},  2503  (1992).

\bibitem{Barenghi96}
C.~F. Barenghi, Phys. Rev. A {\bf 54},  5545  (1996).

\end{thebibliography}
\bibliographystyle{prsty}

\end{document}